\documentstyle[emulateapj]{article}

\received{14 July 1999}
\revised{24 August 1999}

\begin{document}
\title{  TWO DIFFERENT ACCRETION CLASSES IN SEYFERT 1 GALAXIES AND QSOs} 
\author{ Youjun Lu\altaffilmark{1} and Qingjuan Yu\altaffilmark{2}}
\altaffiltext{1}{ Center for Astrophysics, Univ. of Sci. \& Tech. of China, Hefei, Anhui 230026, P.R. China}
\altaffiltext{2}{Princeton University Observatory, Princeton, NJ 08544-1001, USA; Email: yqj@astro.princeton.edu}

\begin{abstract}

The mass of the central black hole in Seyfert galaxies and QSOs can be
determined from the broad emission lines and the reverberation method. 
Using the
measured black hole mass and the bolometric or ionizing luminosity,
the accretion rate can be estimated. Compiling a sample of
Seyfert 1 galaxies and QSOs with reliable central masses, estimated
ionizing luminosities and X-ray spectral slopes, we find that the 
X-ray spectral slope strongly correlates with the accretion rate. The 
objects in the sample are found to be distributed in two distinct classes
in the
spectral index versus Log$(L_{\rm ion}/L_{\rm Edd})$ plane. We argue that
these two classes may correspond to ADAF and thin disk accretion.
The observations of a ``two-state'' Seyfert 1 galaxy, 1H0419-577, confirm our
results. Detailed fitting of the spectra of individual Seyfert 1 galaxies and
QSOs using ADAF and/or thin disk models should further clarify the two-class
classification.
  
\end{abstract}


 
\section{Introduction}

The QSO and active galaxy phenomena can be successfully explained by the
accretion of gas onto central massive black holes (MBHs) (\cite{lyn};
\cite{rees}). Recent progress in both theory and observation strongly
supports this explanation. 
First, compact massive dark objects---possibly
massive black holes---have been detected in the cores of some nearby galaxies
based on high spatial resolution observations of stellar dynamics
(\cite{kr}; \cite{mag}). Second, the radiation
spectra from the nuclei of Sgr A$^*$ and NGC 4258 can be modelled well
by the advection-dominated accretion flows (ADAF) onto MBHs (
\cite{nym}; \cite{las}). In addition, the iron
K$\alpha$ line profile suggests the existence of a MBH and cold disk material
in the center of Seyfert 1 galaxies (\cite{iwa}).

Currently, four solutions of the accretion process around MBHs
are known (\cite{chen}). The most famous two are the thin disk model (\cite{ss})
and the ADAF model (\cite{ich}; \cite{rbbp} Narayan \& Yi
1994, 1995a, 1995b; \cite{abr}), which generally result from
different accretion rate. The radiation spectra output of these models 
are distinctly different (\cite{nmq}). It is therefore reasonably
easy to tell from observations whether a system has a thin or ADAF disk.
The ADAF model provides a good fit to the spectra of Galactic Black
Hole X-ray Binary sources (BHXBs), which should be a low mass accretion version
of Seyfert 1 galaxies and QSOs (\cite{emn}; \cite{esin}). This implies that at
least some Seyfert galaxies and/or QSOs, which were believed to have a thin
disk accretion, may instead accrete material via the ADAF mode. Thus the
question of whether a QSO or an active galaxy has an ADAF or a thin accretion
disk is of great current interest.

To date, the most reliable estimate of masses of central MBHs in
Seyfert 1 galaxies uses reverberation data (\cite{pw}). With
known MBH masses, the accretion rates of active galaxies and QSOs can be
estimated from their bolometric or ionizing luminosity (\cite{wpm},
hereafter WPM). Furthermore, a remarkable correlation
between the soft X-ray spectral indices and the full width at half maximum
(FWHM) of H$\beta$ has been found by Boller, Brandt \& Fink (1996) and Wang,
Brinkmann \& Fink (1996). A possible interpretation is that the correlation is
caused by the variation of accretion rate in different objects(\cite{wy}; 
\cite{pnfm}). The change in soft X-ray spectral slope from
flat to steep may reflect the change in accretion mode from a
low state to a high state. It is reasonable to constrain
theoretical models by investigating the statistical relation between the soft
X-ray spectra and the estimated accretion rates for a sample of objects.

In this letter, we show that a sample of QSOs and Seyfert 1 galaxies with
available estimated MBH masses can be classified into two distinct classes
with different accretion rates. These two classes should correspond to
different accretion disks, i.e. ADAF and thin disk. In each class, the soft
X-ray spectral slope strongly correlates with the estimated accretion
rate. We describe the available data in \S 2. The statistical analysis is
presented in \S 3, and an explanation of the correlation between the X-ray
slope and the accretion rate is discussed in \S 4. 

\section{The Data}

Assuming that the line-emitting matter is gravitationally bound and hence has
a near-Keplerian velocity dispersion (indicated by the line width), WPM
estimated the mass of the central black holes for 17 Seyfert 1 galaxies and 2
QSOs using reverberation data. The central MBH masses were also measured by
the same method for three additional Seyfert 1 galaxies (Mrk279, NGC3516 and
NGC4593) which were not included in WPM's sample (\cite{ho}). Laor (1998)
estimated the MBH masses for 19 QSOs using the
H$\beta$ width and the empirical relation $r_{\rm BLR}({\rm H}\beta)=
15L^{1/2}_{44}$ (\cite{kas}), where $r_{\rm BLR}$ is the size of H$\beta$
emitting region in the broad line region
and ${L_{44}=L(0.1-1\mu{\rm m})}$ in units of
$10^{44}{\rm ergs\ s^{-1}}$. Laor's method may overestimate the black hole
mass, as WPM pointed out that the slope of the empirical relation may be
flatter than 1/2. Fortunately, there is a common object to Laor and WPM's
sample, the QSO PG0953+414, whose mass was estimated to be
$3\times10^{8}{\rm M}_{\odot}$ by Laor, and $1.5\times10^{8}
{\rm M}_{\odot}$ by WPM. We will thus use a calibration factor of 0.5 
for the central black hole masses of the QSOs in Laor's sample. 

Generally, we can estimate the accretion rates for QSOs and Seyfert 1
galaxies from their bolometric luminosity. In WPM's sample, they list the
ionizing luminosity of individual objects, derived from the
reverberation and photoionization method. Following that method, we
estimate the ionizing luminosities for Mrk279, NGC3516 and NGC4593.
Laor (1998) estimated the bolometric luminosities for the QSOs in his sample 
using the empirical relation $L_{\rm bol}=8.3\nu L_{\nu}(3000{\rm \AA})$
(\cite{ld}) for the objects of the sample. For the
common object, PG0953+414, we compare the estimated ${\rm L_{bol}}$ in Laor's
table 1 with ${\rm L_{ion}}$ in WPM, and find that $L_{\rm bol}=3.8L_{\rm ion}$
and $L_{\rm ion}/L_{\rm Edd}\simeq0.53L_{\rm bol}/L_{\rm Edd}$. We take this
relation as an estimation of the Eddington ratio of ionizing luminosity
for the objects in Laor's sample. We list in table 1 the central black hole
masses and the Eddington ratios of ionizing luminosity for all the objects
studied.  Since the ionizing and bolometric luminosities have been estimated,
and the black hole mass determination is the most reliable one, these values
give a direct measurement of the Eddington ratio, $L/L_{\rm Edd}\propto
L_{\rm bol}/M_{\rm BH}$,
and hence the accretion rate $\dot{m}=\dot{M}/\dot{M}_{\rm Edd}$.

In general, the X-ray emission spectra of QSOs and Seyfert 1 galaxies can be
well fitted by a power-law ($f_{\nu}\propto \nu^{-\alpha}$, where $\alpha$ is
spectral index and $\Gamma=1+\alpha$ is photon index). The soft X-ray spectra
for most objects in table 1 are available in the literature. We list the
spectral indices in table 1 both for the soft X-ray range (0.1-2.4keV)
observed by ROSAT and for the hard X-ray range (3-10keV) observed by ASCA. For
the soft X-ray spectra, we adopt the spectral indices of the fitting by using
a power-law with the column density absorption as a free parameter. For
several objects without such fitting, we use the spectral indices determined
assuming a fixed column density absorption as given by the
Galactic column density. There is little difference in the ASCA spectra
fitted by different Fe K$\alpha$ line models. We adopt the photon indices given
by Nandra et al. (1997) in their table 4 by fitting 3-10keV spectra with a
power-law plus a Schwarzschild geometry disk line. The ASCA photon indices of
several objects, which are not included in Nandra et al. (1997), are also
listed in table 1.

\section{Statistical analysis}

Figure 1 shows the spectral indices as a function of the ratio
${\rm Log}(L_{\rm ion}/L_{\rm Edd})$, and hence the accretion rate. We notice
that several objects in our sample, such as 3C273, 3C120, 3C390.3, 3C323.1,
PKS1302-102 and PKS2135-147, are radio-loud galaxies or QSOs.
The X-ray spectra of radio-loud galaxies and QSOs are generally flatter than
those of radio-quiet ones (\cite{lao1}; \cite{wlz}). It 
may be caused by the X-ray emission from the relativistic jet. Indeed, all
the radio-loud objects exhibit flat soft X-ray spectra in our sample. They
will be excluded in the following statistical analysis. In addition, the soft
X-ray spectrum of NGC3227 is very flat, which may be caused by a dusty warm
absorber (\cite{kf}). It will also be excluded in the statistical analysis.

We apply the Pearson linear correlation test (\cite{pre}) to the $\alpha_{
\rm ROSAT}$ versus ${\rm Log}(L_{\rm ion}/L_{\rm Edd})$ relation for
radio-quiet objects. A significant correlation is found, with a linear
correlation coefficient $R=0.80$ ($P_{\rm r}=1.4\times 10^{-7}$).
The soft X-ray spectra become steeper when the Eddington ratio of the ionizing
luminosity increases. It is evident that there is a discontinuity near the
critical point of ${\rm Log}(L_{\rm ion}/L_{\rm Edd})\sim -1.4$ and the data
in figure~\ref{fig-1} are distributed in two separate classes.
One is the lower ${\rm Log}(L_{\rm ion}/L_{\rm Edd})$, representing a lower
accretion rate class, the other is the higher ${\rm Log}(L_{\rm ion}/
L_{\rm Edd})$, representing higher accretion rate class. The two distinct
classes may correspond to the accretion of ADAF and thin disk since the
break point corresponds to the critical threshold between the two
models (see following
section). The Pearson linear correlation coefficients for the two different
classes are $R=0.88$ ($P_{\rm r}=1.4\times 10^{-4}$) and $R
=0.89$ ($P_{\rm r}=2.0\times 10^{-6}$), respectively. They can be fitted by
straight lines:
$\alpha_{\rm ROSAT}=(2.69\pm0.24)+(0.69\pm0.12){\rm Log}(L_{\rm ion}/
L_{\rm Edd})$ for the ADAF class, and
$\alpha_{\rm ROSAT}=(2.00\pm0.06)+(0.60\pm0.08){\rm Log}(L_{\rm ion}/
L_{\rm Edd})$ for the thin disk class. 

Most Seyfert 1 galaxies in the WPM sample belong to the ADAF class and all
QSOs in Laor's sample belong to the thin disk class. It is interesting to
note that five Seyfert 1 galaxies (Mrk110, Mrk335, Mrk509, Mrk590 and NGC4051)
join in the thin disk class. Those Seyfert 1s have small central black holes
but normal Seyfert luminosities. Those support the existence of the two-class
distribution.

The relation between $\Gamma_{\rm ASCA}$ and ${\rm Log}(L_{\rm ion}/
L_{\rm Edd})$ is presented in figure~\ref{fig-2}. There are only 16 objects in
figure~\ref{fig-2} because the others have no ASCA data. We find that there is a
linear correlation between $\Gamma_{\rm ASCA}$ and $L_{\rm ion}/L_{
\rm Edd}$ with a correlation coefficient $R=0.74$ ($P_{\rm r}=0.006$) (after
we exclude the radio-loud objects and NGC3227). The hard X-ray spectra 
become steeper when ${\rm Log}(L_{\rm ion}/L_{\rm Edd})$ increases. 
As figure~\ref{fig-2} shows, the photon indice $\Gamma_{\rm ASCA}$
increases with increasing accretion rate for ADAF class. It
is hard to determine the behavior of the hard X-ray spectra vs. 
${\rm Log}(L_{\rm ion}/L_{\rm Edd})$ for the thin disk class because there
are only 4 objects. Further observations are needed to confirm the
classification.

\section{Discussion}

\begin{figure}
\figurenum{1}
\epsscale{0.5}
\plotfiddle{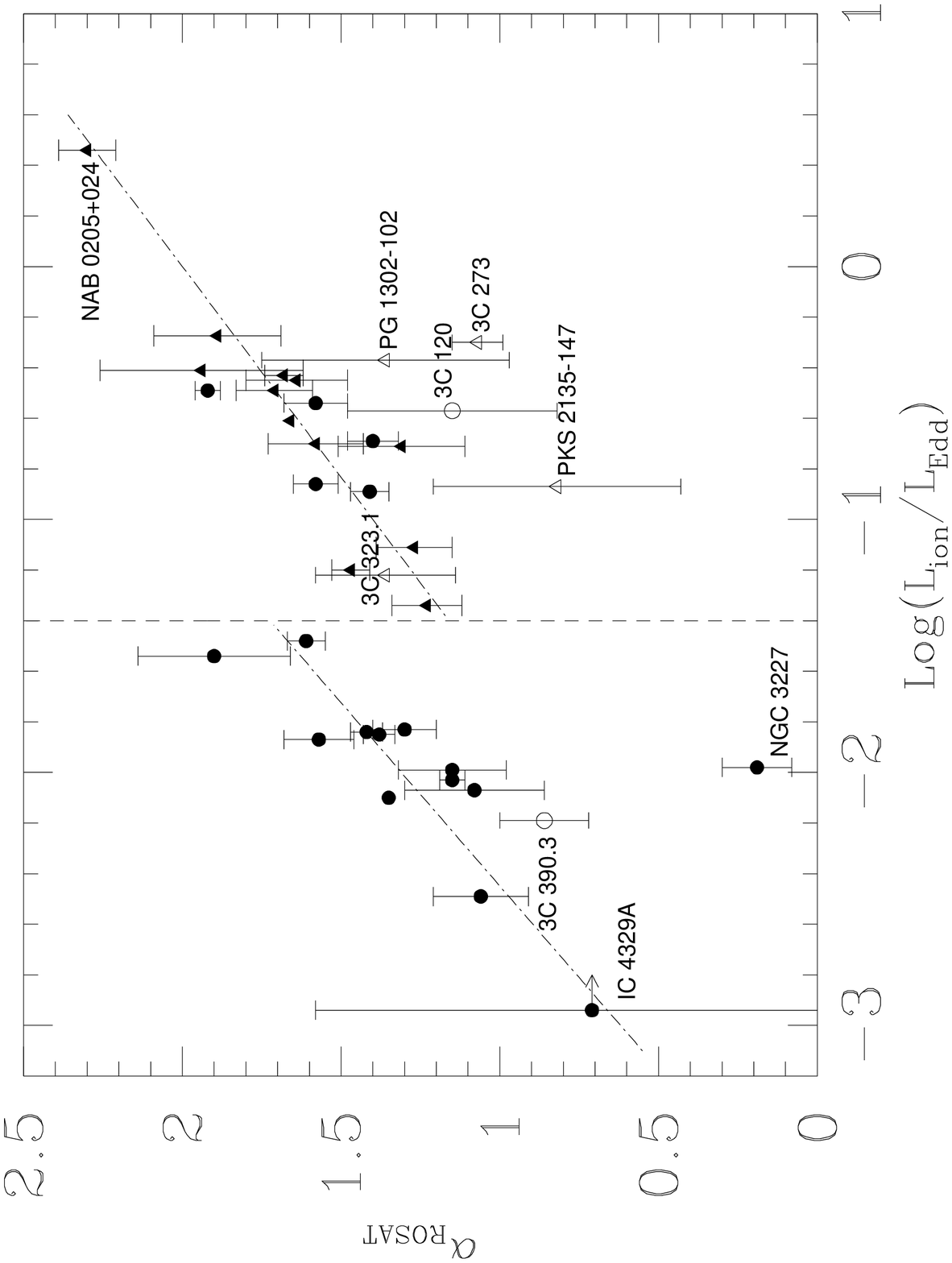}{6.cm}{270}{45}{45}{-200}{250}
\caption[fig1.ps]{The ROSAT soft X-ray spectral indices versus the estimated Eddington ratio of ionizing luminosity. Filled symbols are for radio-quiet
objects; open symbols are for radio-loud objects; circles are for Seyfert 1
galaxies; triangles are for QSOs. The dashed line represents the critical Log
$(L_{\rm ion}/L_{\rm Edd})$ for the ADAF and the thin disk classes.
The dot-dashed lines
are the fitted lines for the two classes. \label{fig-1}}
\end{figure}

\begin{figure}
\figurenum{2}
\epsscale{0.5}
\plotfiddle{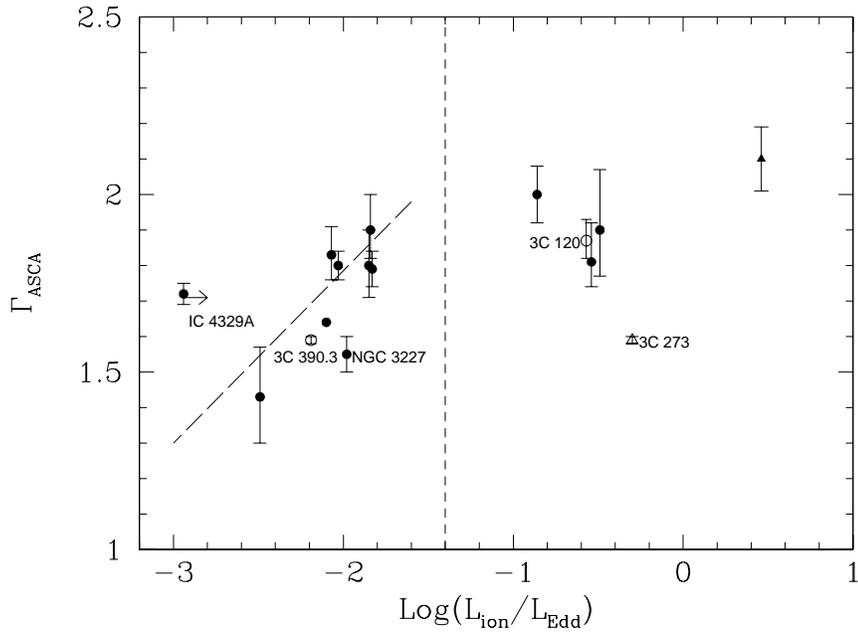}{6.cm}{270}{45}{45}{-200}{250}
\caption[fig2.ps]{The ASCA X-ray photon idices versus the estimated
Eddington ratio of ionizing luminosity. The symbols are the same as in 
figure 1. The dashed line represents the relation of $\Gamma_{\rm ASCA}$
versus Log$(L_{\rm ion}/L_{\rm Edd})$.
\label{fig-2}}
\end{figure}

We find two distinct classes in the ${\alpha_{\rm ROSAT}}$
versus ${\rm Log}(L_{\rm ion}/L_{\rm Edd})$ plane 
for a sample of Seyfert 1 galaxies and QSOs. This classification is striking
because it reveals the existence of two accretion modes in these objects.
The critical threshold between the two classes is ${\rm Log}
(L_{\rm ion}/L_{\rm Edd})\simeq -1.4$ (see in figures~\ref{fig-1} and
figures~\ref{fig-2}). The corresponding ${\rm Log}(L_{\rm bol}/L_{\rm Edd})$ is
about -1.1, which is just the critical accretion rate
${\rm Log}(\dot{m}_{crit})\sim-1.1$ (see the figure 7
in Narayan, Mahadevan \& Quataert 1998). Since $\dot{m}_{crit}\sim \alpha^2$
(\cite{emn}), where $\alpha$ is the viscosity parameter,
the observed discontinuity in figures~\ref{fig-1} suggests $\alpha\sim 0.3$.
Below $\dot{m}_{\rm crit}$, the
accretion is via an ADAF at small radii and a thin disk at large radii.
Because much of the viscously generated energy is advected to the black hole
and $L_{\rm bol}/L_{\rm Edd}\simeq 30\dot{m}^2$ (for the viscosity
parameter $\alpha=0.3$, \cite{emn}), we can derive that $\dot{m}$ of the ADAF
class in our sample is distributed in the range (0.01--0.08), which
corresponds to the low and intermediate states of accretion in BHXBs.
The existence of the ADAF class in Seyfert 1s and QSOs is also supported by the
fact that NGC 4151 has almost an identical spectrum to the BHXB GX339-4 (\cite{zd}).
Above $\dot{m}_{\rm crit}$, the accretion is via a thin
disk and the accretion rate $\dot{m}$ is approximately
$L_{\rm bol}/L_{\rm Edd}$. It may correspond to the high and/or very high
states in BHXBs (\cite{emn}). According to the best fitting straight lines
for the two classes, we derive that $\alpha_{\rm ROSAT}\propto\dot{m}^{1.38}$
for the ADAF class, and $\alpha_{\rm ROSAT}\propto\dot{m}^{0.60}$ for the
thin disk class.

For objects in the thin disk class, it is easy to understand that
the soft X-ray slope steepens with the increase in accretion rate and the 
soft X-ray spectra are steeper than the hard X-ray spectra. The
temperature of the inner thin disk goes up while the accretion rate increases,
therefore the big blue bump moves into EUV and/or soft X-ray range and
the soft X-ray spectrum becomes steeper. As $\dot{m}$ increases, the disk flux
to irradiate the corona increases. This will cause the corona to cool
more efficiently because of Compton cooling, and so the hard X-ray slope will
steepen with the increasing $\dot{m}$. This cannot be confirmed in our sample
because of the lack of ASCA observations for the objects in the thin disk class.
For the ADAF class objects, the X-ray spectra in the ROSAT and ASCA bands both
steepen when $\dot{m}$ increases. This contradicts the ADAF radiation model
of the low accretion state with $\dot{m}\sim (0.01-0.08)$, where the
calculated X-ray spectra should be harder and smoother because the
optical depth goes up and causes a corresponding increase in Compton
y-parameter when $\dot{m}$ increases (\cite{emn}).
However, it can be understood if we adopt a model where the
accretion disk consists of two zones: an outer standard thin disk extending
from a large radius down to $r_{\rm tr}$ and an inner ADAF
from $r_{\rm tr}$ down to $r=3$ (\cite{nar}). If $r_{\rm tr}$ decreases
with increasing $\dot{m}$, and if $r_{\rm tr}$ decreases below
$10^{1.5}$, the ADAF X-ray emission spectra become dramatically softer
because of the cooling effect when the radiation from the disk is
Compton--scattered by the hot gas in the ADAF. It is similar to the
intermediate state in BHXBs (\cite{emn}). The reflection component would
be more important for thin disk and less important
for an ADAF since the solid angle is larger than the former case. This would 
cause the X-ray spectra of the objects in thin disk class softer than those
of the objects the ADAF class (Zdziarski, Lubinski, \& Smith 1999).
We can see in figures~\ref{fig-1} and figures~\ref{fig-2} that the spectral
indices of the objects in the thin disk class tends to be softer than those
of the objects in the ADAF class, statistically.
The accretion flows may have other forms, e.g. ADIOS
(advection dominated inflow and outflow solution, \cite{bb}), which may
be a possible alternative explanation for the relation. Detailed models should be explored to reveal
more information on the two accretion classification.

Recent observations show that 1H0419-577 is a ``two-state'' soft X-ray
Seyfert 1 galaxy (\cite{gun}). In its hard state, the entire
1.8--40keV spectrum can be well-described by a simple flat ($\Gamma \sim 1.55$)
featureless power-law, while in its soft state, the soft X-ray spectrum is
quite steep ($\Gamma \sim 2.5$). The 0.5-2.5keV flux changes by a factor
of $\sim 6$ from its soft state to its hard state. Assuming that $L_{\rm ion}$
is proportional to the flux at 0.5-2.5keV, $L_{\rm ion}/L_{\rm Edd}$ also
changes by a factor of 6. Based on its optical observation, the Eddington
ratio of ionizing luminosity is estimated to be less than 0.08. So, 1H0419-577
is an object in the ADAF class (\cite{gun}). The
Log$(L_{\rm ion}/L_{\rm Edd})$ of this object is predicted to be
$\sim -1.72$ in its soft state by the fitted straight line for the ADAF
class in section 3. Therefore, its Log($L_{\rm ion}/L_{\rm Edd}$) is about
-2.4 in the hard state. The inferred hard X-ray spectral index from
figure~\ref{fig-2}
is consistent with the observations.

We summarize our main result as the existence of two different accretion
classes: the ADAF and the thin disk accretion class in Seyfert 1 galaxies and
QSOs. The soft X-ray spectrum becomes steeper when the accretion rate
increases in both classes. The detailed fitting of Seyfert 1 galaxy and QSO
spectra with ADAF and/or thin disk models should be
explored to reveal the underlying physics, as has been done for BHXBs and 
low-luminosity active nuclei. 

\acknowledgments{We thank an anonymous referee for useful comments which
improve this paper. We thank Neta Bahcall, Scott Tremaine and Joseph 
Weingartner for a critical reading. }


\newpage

\begin{table}
\dummytable\label{tbl-1}
\end{table}

\end{document}